\documentclass[journal=langd5,manuscript=article]{achemso}
\setkeys{acs}{keywords = true}
\usepackage[version=4]{mhchem} 
\usepackage[utf8]{inputenc}
\usepackage[T1]{fontenc}
\usepackage{caption}
\usepackage{graphicx}
\usepackage{xcolor}
\usepackage{ulem}
\usepackage{soul}

\author{Lucie Corral}
\author{Christian Curtil}
\author{Medhi Lagaize}
\author{Marc Leonetti}
\author{Hubert R. Klein}
\email{hubert.klein@univ-amu.fr, marc.leonetti@univ-amu.fr}
\affiliation[CINAM]{Aix-Marseille Univ, CINaM, case 913, campus de Luminy F-13288 Marseille cedex 9, France}

\title{Non-contact mechanics of soft and liquid interfaces by hydrodynamic confinement using a frequency-modulated AFM }


\begin{document}
\maketitle

\begin{abstract}
    Measuring the mechanical response of liquid interfaces without direct contact remains a major experimental challenge, particularly in liquid-liquid systems where no solid reference exists. Here, we develop a frequency-modulation atomic force microscopy (FM-AFM) method that probes liquid interfaces through the hydrodynamic confinement of a viscous liquid film between an oscillating probe and the interface. This approach provides simultaneous access to the in-phase and dissipative components of the effective mechanical response under confinement. The method is first validated on a liquid-solid interface, where the measured confinement thickness and the evolution of the mechanical impedance are consistent with elastohydrodynamic theory over nearly one decade in elastic modulus. It is then applied to a liquid-liquid interface, which exhibits a predominantly viscous response with a finite in-phase contribution and a confinement thickness in the micrometric range. These results show that hydrodynamic confinement provides a sensitive, non-contact approach to compare the mechanical response of soft and liquid interfaces, and opens new perspectives for investigating complex and highly deformable systems such as polymer films, biological membranes, and rafts of nanoparticles.
\end{abstract}

\section{Introduction}

Liquid interfaces play a central role in a wide range of physical, chemical, and biological processes, including wetting, emulsion stability, droplet deformation, and thin-film dynamics~\cite{degennesCapillarityWettingPhenomena2004}. At these interfaces, the mechanical response results from the interplay between viscous flow within the liquid phases and elastic or capillary deformation of the interface~\cite{deGennes1985}. Characterizing this response at small length scales is therefore essential for understanding interactions between fluids and soft matter at micro- and nanometric scales.

Measuring the mechanical properties of soft interfaces is challenging, particularly in the case of liquid-liquid systems. In contrast to solid substrates, a liquid interface cannot sustain a direct mechanical contact: any solid probe brought into contact with the interface inevitably penetrates it. As a result, accessing the mechanical response of such interfaces requires contactless approaches. In this configuration, the measured signal reflects the response of a confined system consisting of the probe, the squeezed liquid film, and the interface.

Over the past decades, interfacial mechanical properties have mainly been investigated using macroscopic or mesoscopic techniques. The Langmuir trough allows the study of monolayers under controlled lateral compression~\cite{Langmuir1917}, while the Wilhelmy plate method provides access to surface tension through force measurements on a partially immersed plate~\cite{wilhelmy}. Dynamic methods such as oscillating or spinning drop techniques probe interfacial tension and viscosity~\cite{hartlandOscillationDropsSpheres1975, freerOscillatingDropBubble2005}. Other approaches, including micropipette aspiration~\cite{hochmuthMicropipetteAspirationLiving2000} or quartz crystal micro balance with dissipation (QCM-D)~\cite{tadros2015viscoelastic}, give access to interfacial viscoelasticity. However, they generally do not allow a clear separation between elastic and dissipative contributions, nor do they operate under strictly contactless conditions.

At smaller length scales, hydrodynamic models based on lubrication theory have been developed to describe the drainage of thin liquid films, and to relate measured forces to viscous dissipation and elastic deformation~\cite{vinogradova1995drainage, vinogradova1999slip, liron1976stokes}. In particular, E. Charlaix group proposed a complete elastohydrodynamic (EHD) description of oscillatory drainage in a thin liquid film confined between a rigid probe and a soft elastic substrate~\cite{leroy2011hydrodynamic}. In this framework, the real and imaginary parts of the mechanical impedance directly reflect elastic deformation and viscous dissipation, providing a quantitative route to separate conservative and dissipative contributions in viscoelastic systems. Frequency-modulated atomic force microscopy (FM-AFM) has also been successfully used to measure hydrodynamic forces in liquids at those scales. As an example, Devailly \textit{et al}~\cite{Devailly_2020} quantified long-range viscous interactions between an oscillating probe and a rigid surface fully immersed in a liquid, demonstrating quantitative agreement with lubrication theory. However, these measurements were restricted to rigid interfaces, for which no deformation occurs and the interaction remains purely hydrodynamic. Extending such approaches to deformable interfaces, and in particular to liquid-liquid interfaces, remains a major experimental challenge.

In the present work, we implement this elastohydrodynamic probing scheme using FM-AFM~\cite{Giessibl2003, albrechtFrequencyModulationDetection1991, garciaDynamicAtomicForce2002}. In FM-AFM, a stiff probe oscillates normal to the interface without direct mechanical contact. The resonance frequency shift $\Delta f$, and the additional dissipation of the probe, provide simultaneous access to the conservative and dissipative components of the interaction with the interface.
The high stiffness and quality factor of the probes ensure sufficient sensitivity in liquid environments while maintaining stable, contactless operation~\cite{ondarccuhu2013nanoscale}.

We first validate this FM-AFM based approach on a reference liquid-solid system, to which the elastohydrodynamic model applies.
This validation step establishes the reliability of the measurement protocol and data analysis for separating elastic and viscous contributions.
We then apply the same experimental method and analysis to a liquid-liquid system.
This configuration constitutes an extreme test case, involving neither solid elasticity nor an established theoretical description.
By extending this hydrodynamic confinement approach from liquid-solid to liquid-liquid systems, this work shows that FM-AFM provides a robust and versatile tool for probing the effective mechanical response of soft and liquid interfaces under confinement. In the liquid-liquid case, the purpose of the present work is primarily to establish the experimental observables and their trends under confinement. A full quantitative theory for mobile liquid-liquid interfaces is left for future work.

\section{Principle of frequency modulated AFM}

In Frequency Modulated Atomic Force Microscopy (FM-AFM), a probe oscillating at a constant amplitude is locked at its resonance frequency. In the vicinity of an interface, the changes in the resonance frequency and in the energy supplied to the oscillator are respectively related to the elastic and viscous components of the probe/interface interaction.

The oscillator (a quartz tuning fork coupled to a glass fiber in our case) is characterized in air, prior to experiments. Its resonance frequency $f_0$ serves as the reference, and is expressed as
\[
f_0 = \frac{1}{2\pi}\sqrt{\frac{k_\text{eff}}{m_\text{eff}}},
\]
where $k_\text{eff}$ and $m_\text{eff}$ are the effective stiffness and mass of the oscillator.
A probe/sample interaction results in changes of the effective stiffness $\Delta k$ and/or mass $\Delta m$, and thus shift the resonance frequency to $f_0'$. We then measure a frequency shift $\Delta f = f_0' - f_0$. Considering small variations in stiffness and mass, it is approximated as:
\[
\frac{\Delta f}{f_0} \simeq \frac{1}{2}\frac{\Delta k}{k_\text{eff}} - \frac{1}{2}\frac{\Delta m}{m_\text{eff}}.
\]

In fluids, for nanometric oscillation amplitudes, $\Delta m$ can be safely neglected, the frequency shift predominantly reflecting the conservative interaction through an interaction-induced stiffness $k_\text{int}$. It is interpreted as the gradient of the conservative part of the probe/sample interaction force. In these conditions, $k_\text{int}$ directly relates to the frequency shift :
\[
k_\text{int} = 2 k_\text{eff} \frac{\Delta f}{f_0} \,
\]
in small oscillation amplitude approximation.

During experiments, the oscillation amplitude $A_0$ of the probe is kept constant. Dissipation is then obtained from the variation of the drive amplitude $A_\text{D}$ of the oscillator. It can be expressed as a normalized damping coefficient\cite{ondarccuhu2013nanoscale,rocheron2022} :
\[
\Delta \beta = \beta - \beta_0 =\beta_0 \left( \frac{A_\text{D}}{A_{\text{D},0}} - 1 \right),
\]
with $\beta_0 = k_\text{eff}/(\omega_0 Q)$ the intrinsic damping of the oscillator, and $A_{\text{D},0}$ its drive amplitude in air.

The harmonic response function of the system, or complex mechanical impedance, is defined as the ratio of the amplitude of the interaction force to the amplitude of oscillation at a given oscillation frequency.
\[
Z(\omega) = \frac{\Tilde{F}_{\omega} }{\Tilde{A}_{\omega}} = Z'(\omega) + i Z''(\omega),
\]
Based on our observables, we can express the moduli of the real and imaginary components of $Z$ at the resonance frequency
\[
Z' = 2 k_\text{eff} \frac{\Delta f}{f_0}, 
\qquad
Z'' = \omega \, \Delta \beta,
\qquad
\omega = 2 \pi f_0.
\]
This formulation allows direct comparisons with dynamic SFA measurements~\cite{leroy2011hydrodynamic}.

\subsection{Probe Design and Calibration}

Our force probe consists of a cylinder glued to a prong of a quartz tuning fork (QTF), and oscillating along its major axis. The quality factors of these QTF oscillators are much higher than for standard AFM probes, which provides force sensitivities as low as a few tenths of pico-newtons per half-hertz. Combined with their high static stiﬀness, this makes them very suitable for operating close to soft interfaces.
Thanks to the hanging fiber design, the oscillator is kept outside the fluid, with only the fiber being dampened. This maintains a high quality factor and therefore high measurement sensitivity.

The cylinders are fabricated from borosilicate glass capillaries tapered using a laser puller (Sutter Instrument Co., P-2000), yielding capillaries with radii of approximately \(2.5~\mu\mathrm{m}\).  

\begin{figure}[H]
    \centering
    \includegraphics[width=0.3\textwidth]{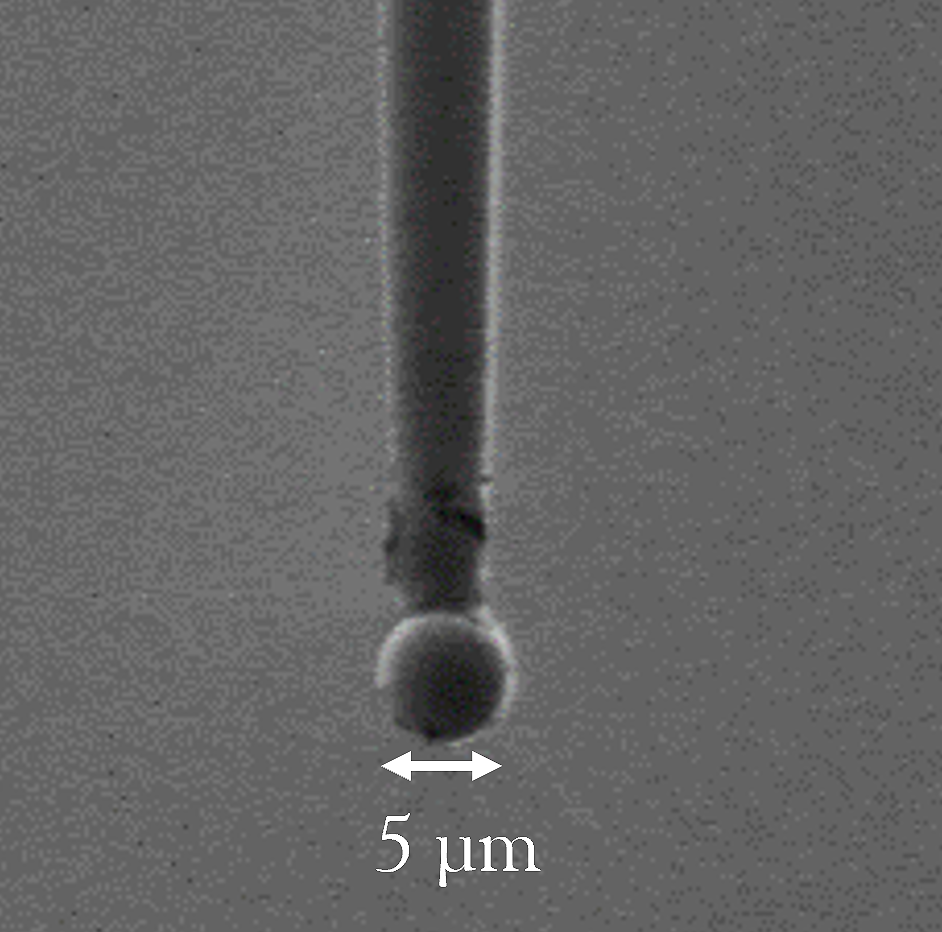}
    \caption{Scanning electron micrograph of the FM-AFM probe. 
    A tapered glass fiber (radius $R = 2.5\mu$m) is attached perpendicularly to one prong of a quartz tuning fork using a UV-curable adhesive (NOA~81). A glass microsphere of 5$\mu$m diameter is glued to the fiber apex, defining a well-controlled sphere-plane geometry suitable for elastohydrodynamic measurements. The tuning fork operates at a resonance frequency $f_0 \simeq 32$kHz, with a quality factor $Q \simeq 10\ 000$, a stiffness $k \simeq 46$kN.m$^{-1}$, and a typical oscillation amplitude $A_0 = 0.5$nm.}
    \label{fig:Probe_SEM}
\end{figure}

A UV-curable adhesive (NOA 81) is used to adhere each fiber perpendicularly to one prong of a quartz tuning fork (QTF), following the probe geometry introduced in ref.~\cite{rocheron2022}. A 5 $\mu$m diameter glass microsphere is affixed to the fiber apex, as shown in figure \ref{fig:Probe_SEM}, in order to define a sphere-plane geometry suitable for controlled hydrodynamic measurements.

A coupled-oscillator model, which takes into consideration the mechanical coupling between the two prongs, is used to determine the tuning fork's stiffness from its geometry and the measured frequencies of the symmetric and antisymmetric oscillation modes~\cite{castellanos2009}, without adjustable parameters. A typical value is $k_\text{eff} = 4.6\pm 0.4\times10^4~\mathrm{N/m}$, with a mean quality factor of $Q=10^4$

We estimate the noise floor~\cite{Butt1995} and determine the electrical sensitivity of our QTFs from measurements of the thermal noise amplitude of the bare oscillator\cite{Welker2011}. Using mean values of the effective stiffness $k_\mathrm{eff}$ and quality factor $Q$, we obtain a thermal oscillation amplitude of the tuning fork $A_\mathrm{th} = 0.3~\mathrm{pm}$ at $T = 295.9~\mathrm{K}$. This corresponds to a force noise density of $1.4~\mathrm{pN\,Hz^{-1/2}}$ and to an electrical displacement sensitivity $\alpha = 65.3 \pm 6.5~\mathrm{nm\,V^{-1}}$. Typical values of probe parameters are shown in Table \ref{tab:probe_params}.

\begin{table}[H]
\centering
\caption{Intrinsic properties of the FM-AFM probe.}
\label{tab:probe_params}
\begin{tabular}{lcr}
\hline
\textbf{Parameter} & \textbf{Symbol / Unit} & \textbf{Value} \\
\hline
Resonance frequency & $f_0$ (kHz) & 32 \\
Oscillation amplitude (free) & $A_0$ (nm) & 0.1 \\
Spring constant & $k$ (N/m) & 4.6$\times$10$^{4}$ \\
Quality factor (air/liquid) & $Q$ & $10^4$ / 3-8$\times10^3$ \\
Tip radius & $R$ (µm) & 2.5\\
Noise floor & $\delta F$ (pN) & $\sim$2 \\
\hline
\end{tabular}
\end{table}

Additional details on probe fabrication, geometrical characterization, calibration procedures and uncertainties, and reproducibility tests are provided in the Supporting Information.

\subsection{Experimental setup}

FM-AFM experiments are performed using a vertical custom-built setup optimized for liquid environments previously described \cite{rocheron2022}.  
The probe is excited by a piezoelectric actuator, and the vertical motion with respect to the interface is controlled using a linear piezoelectric stage.
The entire assembly is enclosed in a sealed chamber with humidity control to minimize evaporation and thermal drift.
Before demodulation, the electrical QTF signal is BP filtered and 20 dB amplified, using a low-noise SRS-560 differential amplifier.
A HF2LI lock-in amplifier demodulates this signal to feed a PLL and a PID feedback loops. PLL loop locks oscillator at resonance frequency by controlling the excitation frequency of the piezo actuator, and monitors the resonance frequency changes ($\Delta f$ channel). The PID loop keeps oscillation amplitude constant by controlling the drive amplitude of the piezo actuator (dissipation channel).
Operating bandwidths of 2 Hz (PLL) and 1 Hz (PID) are used to optimize the signal-to-noise ratios while maintaining stable frequency tracking and amplitude feedback.
 
Experiments are performed in a rectangular reservoir filled with two immiscible phases arranged in a horizontal layered geometry. In the liquid-liquid configuration, the first liquid is introduced into the reservoir, and the second immiscible liquid is carefully deposited on top to form a flat interface. In the liquid-solid configuration, the bottom phase consists of a cross-linked PDMS layer ($\approx1$mm thickness), covered by a liquid layer. 

The FM-AFM probe is positioned above the free surface and vertically displaced across the air-liquid interface, through the upper liquid phase, and toward the liquid-liquid or liquid-solid interface. The thickness of the upper liquid layer is typically $400$-$500~\mu$m. This distance was kept as small as practically possible in order to minimize viscous damping of the fiber and preserve a high quality factor, while remaining sufficiently large to avoid perturbations from the air-liquid interface. As both liquids are transparent, and have similar refractive indices, the interface position cannot be determined by a simple optical method during the experiment. As a result, the probe is approached without direct knowledge of its distance to the interface. The position of the interface is subsequently determined from the hydrodynamic drainage in its vicinity, as described below.

Approach-retract cycles are performed at a constant speed of $50~\mathrm{nm\,s^{-1}}$, much smaller than the vertical oscillation velocity of the sphere \cite{note_v}, ensuring quasi-static measurement conditions. All measurements are conducted at room temperature in PDMS oils or water-glycerol mixtures of controlled viscosity.

For a typical experiment in a $0.24~\mathrm{Pa\,s}$ water-glycerol mixture ($\rho \approx 1230~\mathrm{kg\,m^{-3}}$, $v = 6\times10^{-6}~\mathrm{m\,s^{-1}}$, $R = 2.5~\mu\mathrm{m}$, typical experimental parameters are shown in Table \ref{tab:exp_params}), the Reynolds number is $Re \simeq 2\times10^{-7}$, indicating that inertial effects are negligible and that the flow is in the Stokes regime. The hydrodynamic interaction is thus governed by viscous forces, and the confined flow can be described within the lubrication approximation, corresponding to a Poiseuille-type flow in the thin liquid film. This regime is required for the validity of elastohydrodynamic models used throughout this work.

\begin{table}[H]
\centering
\caption{Experimental parameters used in FM-AFM measurements.}
\label{tab:exp_params}
\begin{tabular}{lcc}
\hline
\textbf{Parameter} & \textbf{Symbol / Unit} & \textbf{Typical Value} \\
\hline
Approach velocity & $v_a$ (nm/s) & 50 \\
Oscillation velocity & $v_o$ ($\mu$m/s) & 6 \\
Reynolds number & $Re$ & $2\times10^{-7}$ \\
Temperature & $T$ (°C) & 20-25 \\
\hline
\end{tabular}
\end{table}

\subsection{Mechanical stability of the fiber probe.}

It is necessary to check the mechanical stability of the cantilevered fiber probe before frequency and dissipation shifts can be interpreted. Because of its high aspect ratio, the glass fiber may undergo lateral deflections under axial loading. The probe would experience lateral deflections as a result of this coupling, leading to frequency and dissipation shifts unrelated to the intrinsic interfacial mechanical response.

The onset of buckling is estimated using Euler’s column theory~\cite{timoshenko1936theory}. Since the probe consists of a hollow cylindrical glass capillary, its bending stiffness is characterized by the second moment of area
\[
I = \frac{\pi}{64} \left( d_o^4 - d_i^4 \right),
\]
where $d_o = 5~\mu$m is the outer diameter and $d_i \simeq 1.8~\mu$m the inner diameter of the fiber. The corresponding critical buckling load is
\[
P_\mathrm{cr} = \frac{\pi^2 E_f I}{(K L)^2},
\]
with $E_f \approx 70$~GPa is the Young’s modulus of borosilicate glass, $L = 3$~mm the effective length of the fiber, and $K = 2$ the boundary condition factor for a fixed-free geometry. This yields a critical buckling force $P_\mathrm{cr} \simeq 5.8 \times 10^{-7}~\mathrm{N}$. 
To relate this axial load to the experimental indentation depth, we use Hertzian contact mechanics~\cite{johnson1985contact} for a sphere of radius $R = 2.5~\mu$m indenting an elastic substrate:
\[
P = \frac{4}{3} E^* \sqrt{R} \, \delta^{3/2}, \qquad
E^* = \frac{E_s}{1 - \nu^2},
\]
where $E_s$ is the Young’s modulus of the substrate and for PDMS, we assume a nearly incompressible behavior with a Poisson ratio $\nu = 0.48$, as experimentally reported for crosslinked PDMS.\cite{muller2019quick}. Solving for $\delta$ at $P = P_{cr}$ gives the critical indentation depth
\[
\delta_{cr} = \left( \frac{3 P_{cr}}{4 E^* \sqrt{R}} \right)^{2/3}.
\]

For the stiffest PDMS investigated ($E_s = 2.7$~MPa), this yields $\delta_{cr} \simeq 180$nm.

Since FM-AFM directly measures the interaction stiffness $Z' = dP/d\delta$, it is useful to express this limit in terms of a critical stiffness. Differentiating the Hertz relation gives
\[
Z = \frac{dP}{d\delta} = 2 E^* \sqrt{R \delta}.
\]

Evaluating at $\delta = \delta_{cr}$ yields the critical interaction stiffness

\[
Z_{cr} =
2 E^* \sqrt{R \delta_{cr}}
=
\frac{3}{2} \frac{P_{cr}}{\delta_{cr}}
\approx 4.8~\mathrm{N.m^{-1}}
\]

This value defines the maximum interaction stiffness that can be applied while maintaining axial stability of the fiber probe. For $Z < Z_{cr}$, the probe remains mechanically stable and the measured impedance reflects the interfacial response. For $Z$ approaching $Z_{cr}$, probe compliance and the onset of buckling may affect the measurements, and data points approaching this threshold were excluded from the quantitative analysis.
Using this Euler-Hertz estimate, the corresponding critical indentation depth for stable operation lies in the range $\delta_\mathrm{cr} \approx 180\text{-}200~\mathrm{nm}$.
For the stiffest PDMS substrate ($E_s = 2.7$~MPa), the measured interaction stiffness approached this threshold, indicating that probe mechanics rather than interfacial mechanics limited the accessible range. In contrast, for softer substrates ($E_s = 0.4$-$1.2$~MPa), the interaction stiffness remained well below $Z_{cr}$, and probe stability is not a limiting factor.
This constraint does not apply to liquid-liquid interfaces, where the absence of bulk elasticity prevents significant axial compression of the probe.

\section{Results and discussion}

\subsection{FM-AFM Observables}

During approach-retract cycles, the normalized frequency shift $\Delta f/f_0$ and the excess dissipation are recorded as a function of the probe vertical position. These signals provide access, in the vicinity of the interface, to the conservative and dissipative components of the sphere-interface interaction mediated by the confined liquid film.

In the bulk hydrodynamic regime, where the interface does not affect the probe motion, both signals vary linearly with the immersion depth due to viscous drag. This regime corresponds to hydrodynamic loading by the bulk liquid, whose magnitude depends on the liquid viscosity, density and varies linearly with the penetration depth~\cite{rocheron2022}. Other studies have focused on the rheology of simple fluids \cite{Sandoval2015, Xiong2009}, but here we are interested in the response associated with confinement near the interface.
Thus, to isolate the contribution associated with confinement, the hydrodynamic linear background was subtracted from both signals. 
The corrected data were then converted into the real and imaginary components of the complex mechanical impedance, $Z = Z' + iZ''$. In the following, $Z'$ denotes the in-phase component of the measured response, while $Z''$ denotes its dissipative component. In liquid-solid systems, these quantities can be directly interpreted within the elastohydrodynamic framework. In liquid-liquid systems, however, they should be regarded more cautiously as effective quantities describing the response of the coupled probe-film-interface system under hydrodynamic confinement.

\subsection{Reference Liquid-Solid Interface}

We first apply the FM-AFM hydrodynamic probing scheme to a reference liquid-solid system consisting of a water-glycerol mixture confined against a cross-linked PDMS thick film ($E=2.7$MPa). In this configuration, the elastohydrodynamic (EHD) framework provides a quantitative description of the oscillatory drainage of a thin liquid film of thickness $D$ confined between the rigid sphere and an elastic interface. In this regime, the hydrodynamic pressure remains approximately constant within a zone of radial extension $\sim\sqrt{RD}$, resulting in a uniform elastic deformation of the substrate.

\begin{figure}[H]
    \centering
    \includegraphics[width=0.7\textwidth]{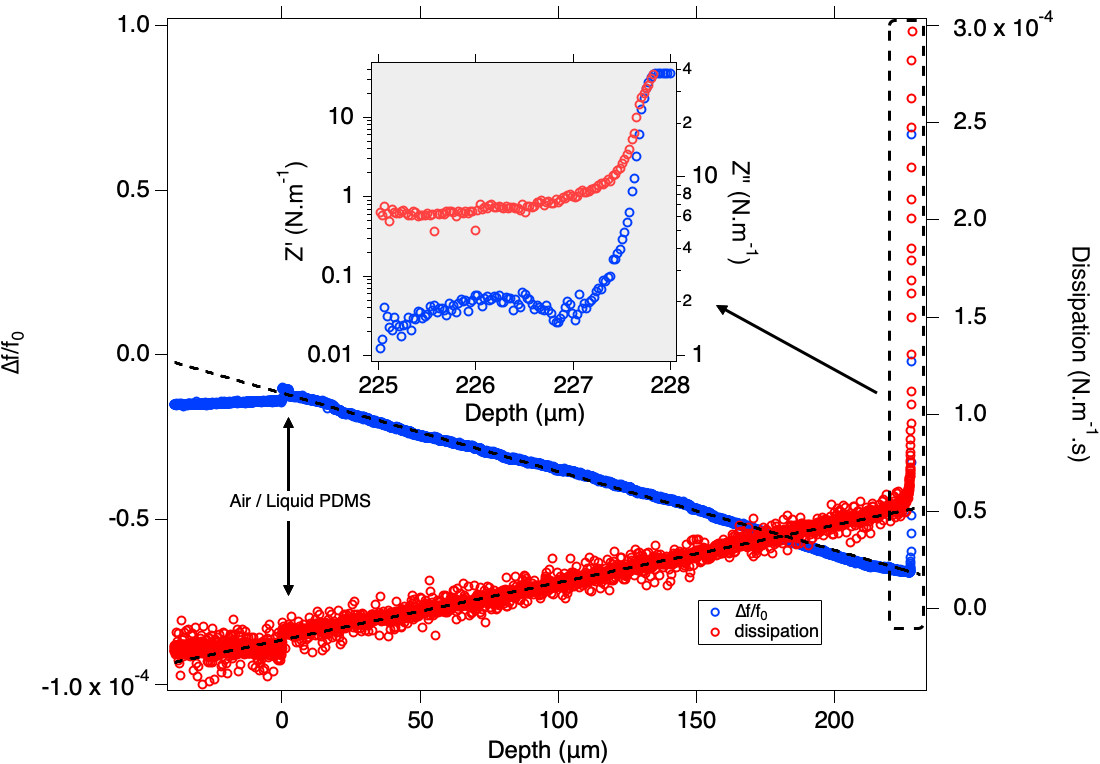}
    \caption{Raw AFM signals recorded during the approach of the oscillating probe toward a cross-linked PDMS thick film ($E=2.7$MPa) in a water-glycerol mixture. The normalized resonance frequency shift $\Delta f/f_0$ (blue triangles, left axis) and the dissipation signal (red circles, right axis) are plotted as a function of probe depth. The origin of depth has been set at the position of the air/PDMS interface. After entering the liquid (vertical arrows), both observables exhibit a linear dependence on depth (dashed lines), characteristic of viscous drag in the bulk liquid, followed by a transition towards interfacial confinement and deformation. After subtraction of the bulk liquid contribution, the inset shows the real ($Z'$, blue) and imaginary ($Z''$, red) components of the mechanical impedance as a function of depth.}
    \label{fig:RawData_PDMS}
\end{figure}

Figure~\ref{fig:RawData_PDMS} shows the raw signals recorded during the approach. Both the normalized resonance frequency shift $\Delta f/f_0$ and the dissipation signal vary linearly with immersion depth in the bulk liquid, reflecting viscous drag. After subtraction of this contribution, the real and imaginary parts of the mechanical impedance, $Z'$ and $Z''$, are calculated. In this raw representation, the probe depth is measured relative to the air / liquid PDMS interface. The position of the undeformed interface is determined later from the extrapolation of the linear hydrodynamic regime of $1/Z''$, as described below.

In the hydrodynamic regime, where no elastic deformation occurs, the interaction is governed by Reynolds lubrication theory. In the sphere-plane geometry, it gives an expression for the dissipative impedance \cite{vinogradova1995drainage, Israelachvili2011}:

\[
Z''(\omega) = \frac{6\pi \eta R^2 \omega}{D},
\]

where $D$ is the distance to the undeformed interface. Consequently, $1/Z''$, the mobility, varies linearly with $D$ and can be used to determine the absolute interface position by extrapolation to zero mobility, known as the hydrodynamic zero~\cite{canaleNanorheologyInterfacialWater2019}.

\begin{figure}[H]
	\centering
	\includegraphics[width=0.7\textwidth]{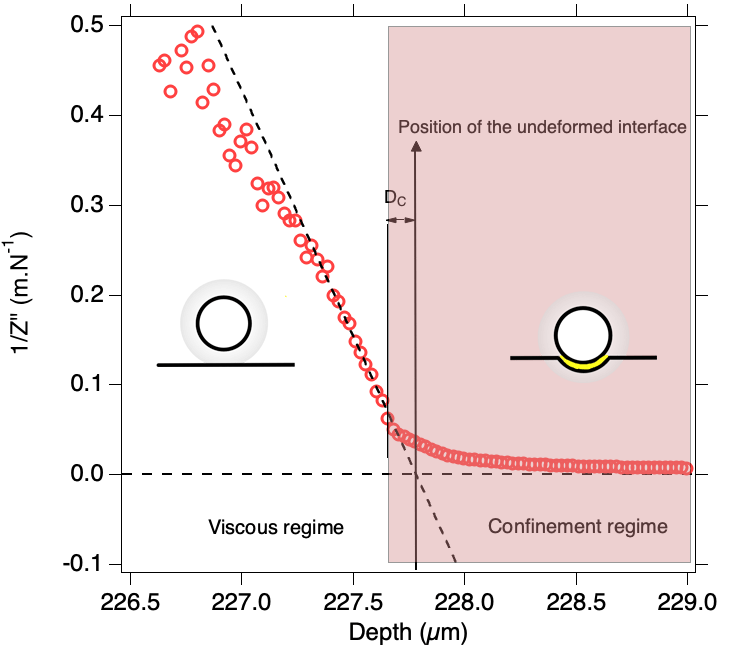}
	\caption{Determination of the confinement thickness $D_c$ at the water-glycerol / cross-linked PDMS interface ($E = 2.7$ MPa, $\eta = 0.013$ Pa·s, $T = 22^\circ$C). The inverse dissipative impedance, $1/Z''$ (mobility), is plotted as a function of probe depth. Far from the interface, $1/Z''$ follows a linear dependence characteristic of viscous drainage in the lubrication regime. The linear fit (dashed line) defines the hydrodynamic regime. The extrapolation of this regime to zero mobility determines the position of the undeformed interface. At shorter distances, the signal deviates from this linear behavior and reaches a plateau, corresponding to the confinement regime where viscous drainage is hindered by interfacial deformation. The shaded region and schematic representations indicate the two regimes: a far-field viscous regime (no deformation) and a confined regime where the interface deforms under hydrodynamic stress. The confinement thickness is defined as the distance between the undeformed interface and the onset of deviation from the viscous regime, yielding $D_c = 135 \pm 20$ nm.}
	\label{fig:invZ_PDMS}
\end{figure}

Figure~\ref{fig:invZ_PDMS} shows such an analysis of the evolution of $Z''$ on cross-linked PDMS, which allows us to determine the initial position of the interface.

 We also determine the confinement thickness from the distance with respect to the position of the undeformed interface, at which 1/Z" deviates from a dominantly viscous behavior with a capillary contribution. This analysis yields $D_c = 135 \pm 20~\mathrm{nm}$, in quantitative agreement with the elastohydrodynamic prediction,

\[
D_c = 8R \left( \frac{\eta \omega}{E^*} \right)^{2/3},
\]

which gives $D_c = 154~\mathrm{nm}$ for the present experimental parameters. The uncertainty associated with this procedure, and measurements reproducibility are discussed in the Supporting Information.

\begin{figure}[H]
    \centering
    \includegraphics[width=0.7\textwidth]{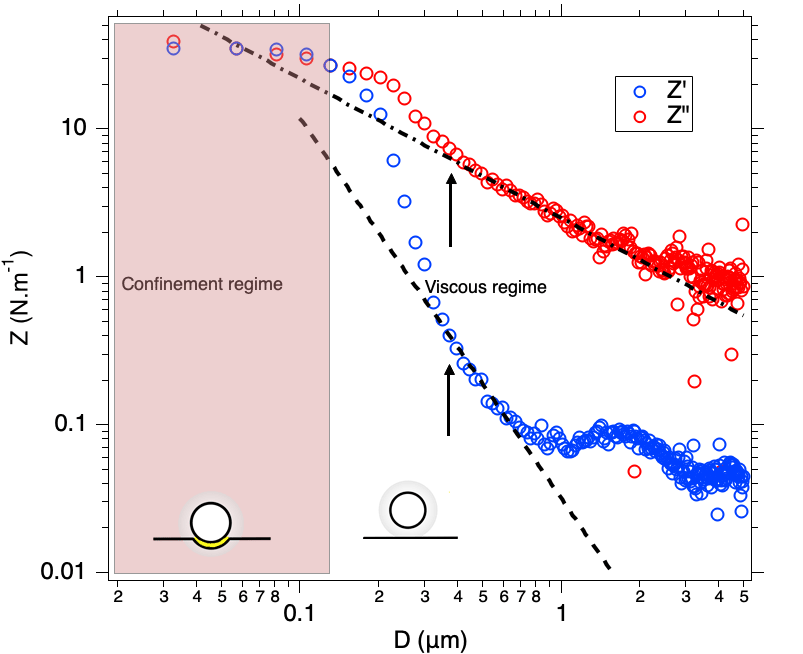}
    \caption{Plot of the real and imaginary parts of the mechanical impedance, at the water-glycerol cross-linked PDMS interface ($E = 2.7$MPa, $\eta = 0.013$Pa.s, $T = 22^{\circ}$C), as a function of the probe-undeformed interface distance. Showing, as the probe moves closer to the interface, the transition from viscous behaviour to elastic deformation, where both components saturate at similar values, as expected for a bulk viscoelastic material. In the elastic deformation regime before saturation, $Z"$ follows a $D^{-0.94\pm0.03}$ power law, while $Z'$ follows a $D^{-2.6\pm 0.1}$ law. The arrows indicate the distance beyond which fiber buckling occurs (see text for details). The shaded areas and schematic illustrations emphasize the transition between viscous drainage (no deformation) and elastohydrodynamic coupling (deformation of the substrate).}
    \label{fig:Z_PDMS_Dc}
\end{figure}

Using this reference position, the impedance components can be plotted as a function of the distance to the undeformed interface (Fig.~\ref{fig:Z_PDMS_Dc}). In this representation, we clearly distinguish two regimes. Far from the interface, the dissipative component follows the Reynolds scaling $Z'' \propto D^{-1}$, with a measured exponent of $-0.94 \pm 0.03$. Simultaneously, the elastic component follows the elastohydrodynamic prediction $Z' \propto D^{-5/2}$, with a measured exponent of $-2.6 \pm 0.1$, in quantitative agreement with the theoretical value. 
At approximately $300$nm from the interface (indicated by arrows in the figure \ref{fig:Z_PDMS_Dc}), the impedance exceeds the critical value defined above, and fiber buckling occurs. However, this did not prevent us from determining the elastohydrodynamic scaling and does not affect the quantitative agreement with the model. 

As the probe approaches closer than $D_c$, viscous drainage is inhibited and the hydrodynamic pressure is transmitted to the elastic substrate, which deforms in response to the oscillatory loading. In this regime, both impedance components saturate to comparable values.

\begin{figure}[H]
	\centering
	\includegraphics[width=0.6\linewidth]{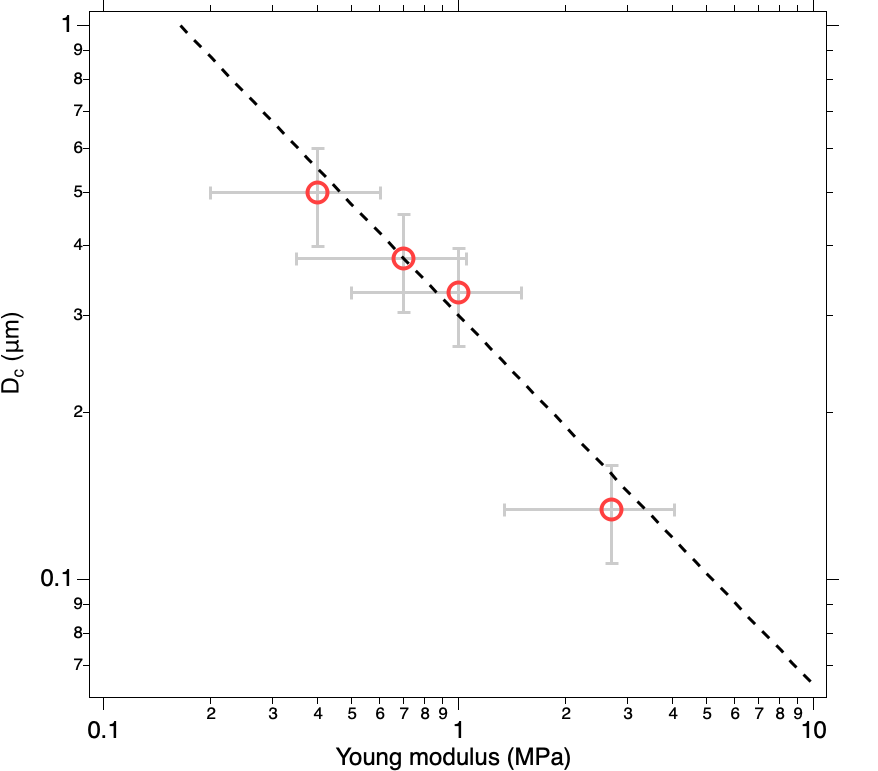}
	\caption{Confinement thickness $D_c$ measured at the water-glycerol / cross-linked PDMS interface as a function of the Young’s modulus $E$ of the elastomer (symbols). The dashed line shows the elastohydrodynamic prediction $D_c = 8R(\eta \omega / E^*)^{2/3}$. Experimental uncertainty on $D_c$ arises from the resolution of the positioning stage and from the determination of the hydrodynamic zero (see supplementary information for details). The agreement with the theoretical trend confirms the quantitative reliability of the method.}
	\label{fig:Dc_vs_E}
\end{figure}

To further validate this elastohydrodynamic description, we then vary the Young’s modulus of the cross-linked PDMS and reproduce the experiment. The resulting measured confinement thicknesses $D_c(E)$ are plotted in Fig.\ref{fig:Dc_vs_E}.

The measured confinement thickness follows the predicted scaling $D_c \propto E^{-2/3}$, in quantitative agreement with elastohydrodynamic model. This liquid-solid reference system therefore provides a quantitative validation of the FM-AFM hydrodynamic probing method.

\subsection{Liquid-Liquid Interface}

The FM-AFM measurements are then performed on a liquid-liquid interface composed of two immiscible viscous phases: a PDMS oil phase ($\eta_{\mathrm{PDMS}} = 0.019$Pa.s) and a water-glycerol mixture ($\eta_{\mathrm{aq}} = 0.246$Pa.s). 
Under these conditions, the viscous diffusion length associated with the probe’s oscillation in PDMS is $l=\sqrt{\eta / \rho f} \simeq 20\mu$m for an oscillation frequency of 32 kHz, allows contactless interaction under these conditions.
We have deliberately chosen a highly viscous underlying aqueous phase to slow down drainage and enhance the hydrodynamic stresses transmitted to the interface. We therefore expect to improve the sensitivity to the mechanical response, approaching conditions comparable to a liquid-solid interface. 

\begin{figure}[H]
	\centering
	\includegraphics[width=0.7\linewidth]{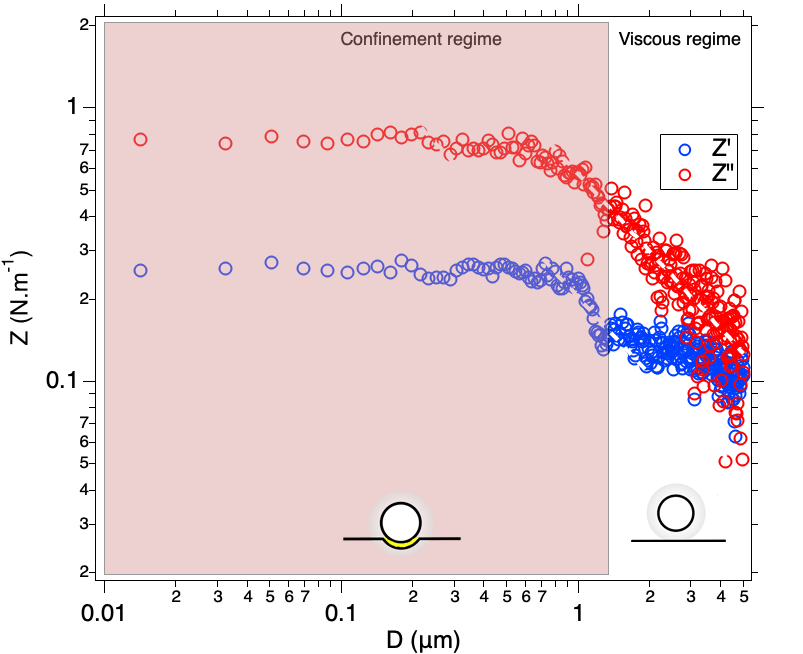}
	\caption{Real ($Z'$) and imaginary ($Z''$) components of the mechanical impedance measured by FM-AFM during the approach toward a fully liquid-liquid interface composed of PDMS ($\eta_{PDMS} = 0.019$Pa.s) and a water-glycerol mixture ($\eta_{aq} = 0.246$Pa.s), at $T = 22~^{\circ}$C. The probe approached quasi-statically at $v_{app} = 50$nm.$s^{-1}$ with an oscillation velocity $v_{osc} = 6 \mu$m.$s^{-1}$. The impedance components are plotted as a function of the probe-interface separation for a glass microsphere of radius $R = 2.5 \mu$m. 
	While approaching the interface, a transition is observed between a distance-dependent viscous regime and a saturation regime, which indicates interaction via a confined fluid film (highlighted by shaded region). In contrast to the liquid-solid case, $Z'$ saturate to a lower value than $Z"$, suggesting a predominantly viscous response dominated by hydrodynamic stresses. 
	Both $Z'$ and $Z''$ remain more than one order of magnitude smaller than in the liquid-solid case, reflecting the absence of bulk solid elasticity and a predominantly viscous response governed by hydrodynamic drainage and interfacial deformation.}
	\label{fig:Z_LL}
\end{figure}
Figure \ref{fig:Z_LL} shows the real and imaginary components of the mechanical impedance measured during the approach toward the liquid-liquid interface. As in the liquid-solid configuration, a transition is observed between a distance-dependent hydrodynamic regime at large separations and a plateau at short distances. This plateau defines an effective confinement thickness associated with the onset of interfacial deformation, corresponding to the transition between a far-field hydrodynamic drainage regime and a short-distance regime where the interface contributes significantly to the measured response.

Compared to the liquid-solid reference system, both impedance components are reduced by more than one order of magnitude. In the absence of a bulk solid elasticity, the mechanical impedance exhibits a measurable in-phase component response, the saturation value of which remains below that of $Z"$. We measure a change in Z' relative to the interface deformation of $108 \pm 20~\mathrm{mN\,m^{-1}}$, which is greater whilst remaining of the same order of magnitude than the intefacial tension ($\gamma = 37 \pm 1~\mathrm{mN\,m^{-1}}$ measured by the pending drop method \cite{rocheron2022}). This suggests that the flow at the liquid interface differs from that described at the liquid solid interface as expected from a liquid-liquid interface.

We extract the confinement thickness from the inverse of the dissipative impedance, $1/Z''$, following the procedure described above. A value of $D_c = 1.5 \pm 0.3~\mu\mathrm{m}$ is obtained, approximately one order of magnitude larger than the one measured for the liquid-solid interface. 

The reproducibility of this result was assessed using repeated measurements performed with different probes. Consistent confinement thicknesses were obtained across experiments, as detailed in the Supporting Information.  This result indicates that with this setup, we are able to induce deformation and measure the response of the interface, even in the absence of a bulk elasticity, and confirms the much higher deformability of a liquid interface compared to elastic solids.

\subsection{Discussion}

\begin{figure}[H]
    \centering
    \includegraphics[width=0.6\textwidth]{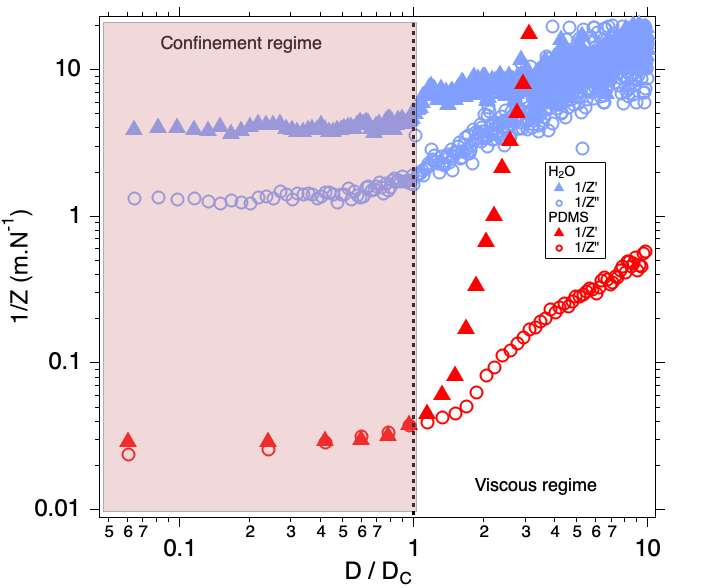}
     \caption{Real (solid triangles) and imaginary (open circles) components of the inverse mechanical impedance ($1/Z'$ and $1/Z''$) as a function of the normalized distance to the interface for a liquid-solid system (water-cross-linked PDMS, red) and a liquid-liquid system (PDMS oil-water, blue). In both systems, the dissipative component varies linearly with distance in the far-field regime (dotted lines), consistent with viscous drainage.  
     In the liquid-solid case, the response reflects elastohydrodynamic coupling with a viscoelastic substrate, leading to a large in-phase component. In contrast, the liquid-liquid interface exhibits a much weaker response, with both components remaining significantly smaller and dominated by viscous effects.
     The confinement thickness is identified at $D_c = 135 \pm 20$ nm for the liquid-solid interface and $D_c = 1.5 \pm 0.3~\mu$m for the liquid-liquid interface.
     The shaded regions indicate the confinement regime ($D < D_c$), while the dashed vertical line marks the transition from viscous drainage to interfacial deformation.}
    \label{fig:liq_liq_combined}
\end{figure}

The most severe test of the sensitivity of the FM-AFM hydrodynamic probing method is provided by the liquid-liquid interface, where both contacting phases are deformable and no bulk elasticity contributes to the mechanical response. In contrast with liquid-solid systems, where the no-slip boundary condition imposes zero velocity at the interface and leads to classical Poiseuille drainage, a liquid-liquid interface allows interfacial motion. As a result, the flow field and stress distribution differ fundamentally from the elastohydrodynamic framework. Figure \ref{fig:liq_liq_combined}, which shows the plot of the inverse impedances as a function of the reduced distance $D/D_c$ from the interface for the two systems studied, illustrates this difference.

In the liquid-liquid configuration, the measured mechanical impedance should therefore not be interpreted as an intrinsic material property of the interface alone. Instead, it reflects the response of a coupled system composed of the probe embedded in the oil, the confined viscous film, the deformable interface and the underneath liquid, the water solution. As a result, the extracted impedance is more appropriately viewed as an effective response under hydrodynamic confinement.

The finite in-phase response observed at the liquid-liquid interface cannot be attributed solely to static capillary forces. Although $Z'_{\mathrm{sat}}$ exceeds the equilibrium interfacial $\gamma$, the observed scaling and the comparable magnitude of $Z'$ and $Z''$ indicate that the response is governed by viscous stresses and interfacial flow rather than by an elastic restoring mechanism. More specifically, this in-phase component can be interpreted as arising from phase-shifted hydrodynamic stresses generated by the deformation and motion of the interface under oscillatory confinement. It therefore reflects an effective dynamical response of the coupled system rather than a true elastic modulus.

Quantitatively, the confinement thickness increases from $D_c \simeq 0.1\mu$m in the liquid-solid case to $D_c \simeq 1.3\mu$m in the liquid-liquid configuration. This order-of-magnitude increase directly reflects the absence of bulk elastic restoring forces and the enhanced deformability of the interface.

This also highlights the fundamentally different nature of the mechanical response in liquid--liquid systems, where deformation is governed by the  flow rather than elastic compression as expected for a liquid/solid interface.
The EHD model cannot be directly applied to this configuration. It assumes that the interfacial velocity is zero in the absence of slip. While this condition is satisfied for a liquid/solid interface, for a liquid/liquid interface the interfacial velocity is continuous through the interface without slip in the case of two no miscible liquids of small molecular size leading to a difference with the Poiseuille-type flow. If the chain length of PDMS is increased to change the viscosity contrast, slippage cannot simply be eliminated without a quantitative analysis.

A quantitative description of this configuration would require accounting for the velocity continuity and mechanical equilibrium at the liquid-liquid interface, including the viscosity contrast between the two phases.  as long as we are considering these kinds of liquids, slippage should not play a decisive role. At the leading order, we expect that the velocity continuity leads to an hydrodynamic torus loop around the axis of symmetry, whose upward central velocity makes the interface stiffer. However, only a careful comparison between a model and the experimental results on a large range of viscosity contrast will be conclusive.

These results demonstrate that, even in the absence of bulk elasticity, the weak mechanical response of liquid interfaces can be measured by this technique. They show that this response under hydrodynamic confinement is governed by viscous stresses and interfacial flow and by elastic restoring forces.

\section{Conclusion}

We have developed a frequency-modulation AFM method to probe liquid interfaces under hydrodynamic confinement in a strictly contactless manner. Using a hanging fiber probe, both the dissipative and in-phase components of the confined mechanical response, as well as an effective confinement thickness, can be determined with high sensitivity.
The method was first validated on a reference liquid-solid interface, where the measured evolution of the impedance and the confinement thickness are consistent with the elastohydrodynamic model. It was then applied to a fully liquid-liquid interface, where no bulk elasticity is present.
The measurements reveal a predominantly viscous response together with a finite in-phase contribution in the liquid-liquid case. These results show that hydrodynamic confinement provides access to the coupled dynamics of confined liquid films and deformable interfaces. In this configuration, the measured response should be interpreted as an effective response of the confined system rather than as a direct determination of intrinsic interfacial properties.
This distinction is essential when interpreting measurements on liquid–liquid systems, where the absence of bulk elasticity prevents a direct identification of intrinsic interfacial mechanical properties.
More broadly, this approach establishes a new experimental framework for comparing the micromechanical response of soft and liquid interfaces in a fully contactless geometry. Even though the current implementation does not allow for spectroscopic measurements, it nevertheless opens promising perspectives, in addition to SFA-type techniques, for studying at a local scale complex and highly deformable systems such as polymer films, biological membranes, and other interfaces where confinement and interfacial flow play a central role.

\section*{Supplementary informations}

\subsection*{Probe Geometry and Fabrication}

The probe used in this study consists of a glass capillary attached to one prong of a quartz tuning fork (QTF), and terminated by a glass microsphere defining a sphere-plane geometry.

\subsubsection*{Preparation of the tuning fork}

The quartz tuning forks are initially encapsulated in a sealed metallic casing. In order to access the prongs for probe assembly, the end of this casing is carefully removed using a watchmaker’s lathe. This operation preserves the mechanical and electrical integrity of the tuning fork while allowing direct access to the prongs.

\subsubsection*{Capillary fabrication}

The capillaries are fabricated from borosilicate glass tubing (outer diameter 1 mm, inner diameter 0.5 mm, length 10 cm, item \#B100-50-10). They are stretched using a laser puller, resulting in a thinned region at the center of the capillary.

At this stretched region, the capillary is separated into two parts. In this region, the outer diameter is typically $\sim 5~\mu$m and the inner diameter $\sim 3~\mu$m over a length of several centimeters. This process ensures a smooth and continuous reduction of the capillary dimensions.

The extremity resulting from the pulling process defines the probe apex. This method provides a clean and well-defined tip geometry, in contrast to mechanical cutting, which may introduce irregularities or angular defects.

Prior to assembly, the capillaries are characterized using electron microscopy to verify their diameter, uniformity, and absence of surface defects.

The apex of the capillary exhibits a slight constriction over a few microns, corresponding to a partial pinching of the inner channel. This feature does not affect the measurements, as the capillary is not used for fluid transport. In particular, this geometry does not introduce any significant modification of the hydrodynamic flow at the length scales probed, which remain much larger than the size of this region. Away from this zone, the capillary retains a regular cylindrical geometry.

\begin{figure}[ht]
	\centering
	\includegraphics[width=0.45\textwidth]{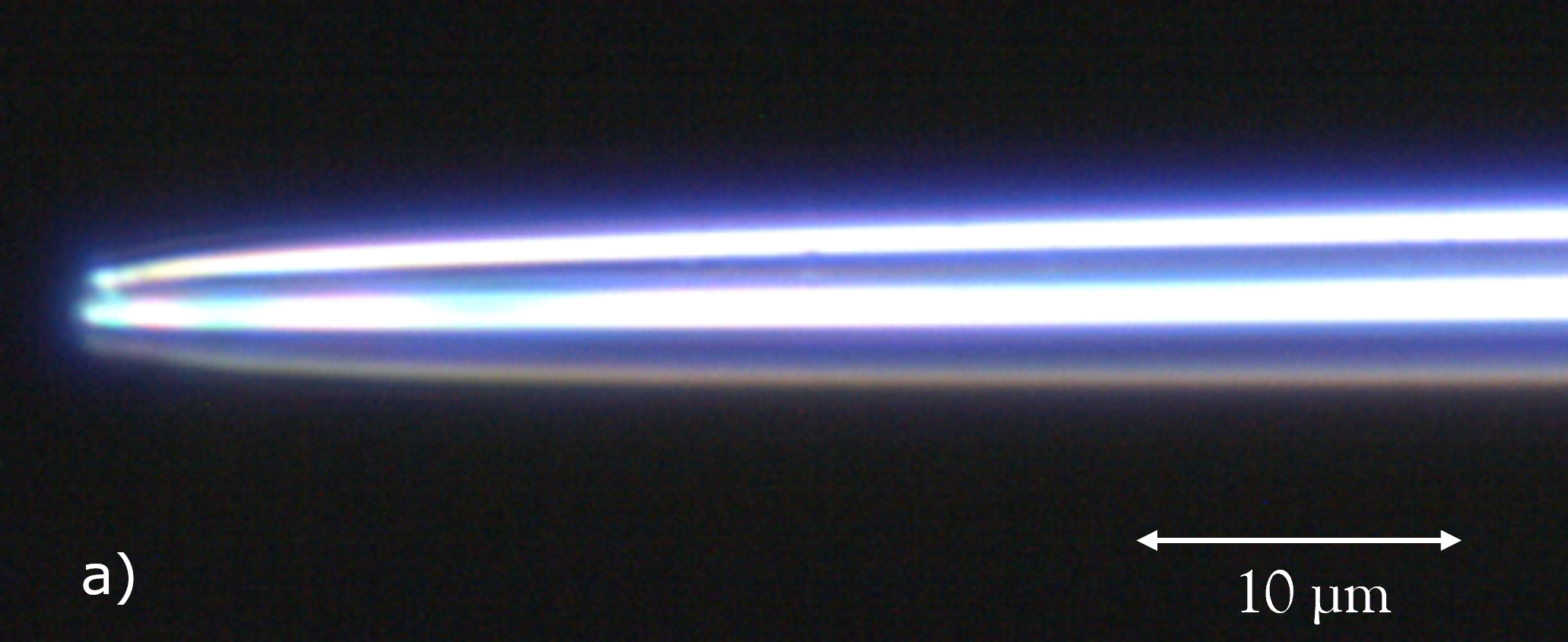}
	\includegraphics[width=0.45\textwidth]{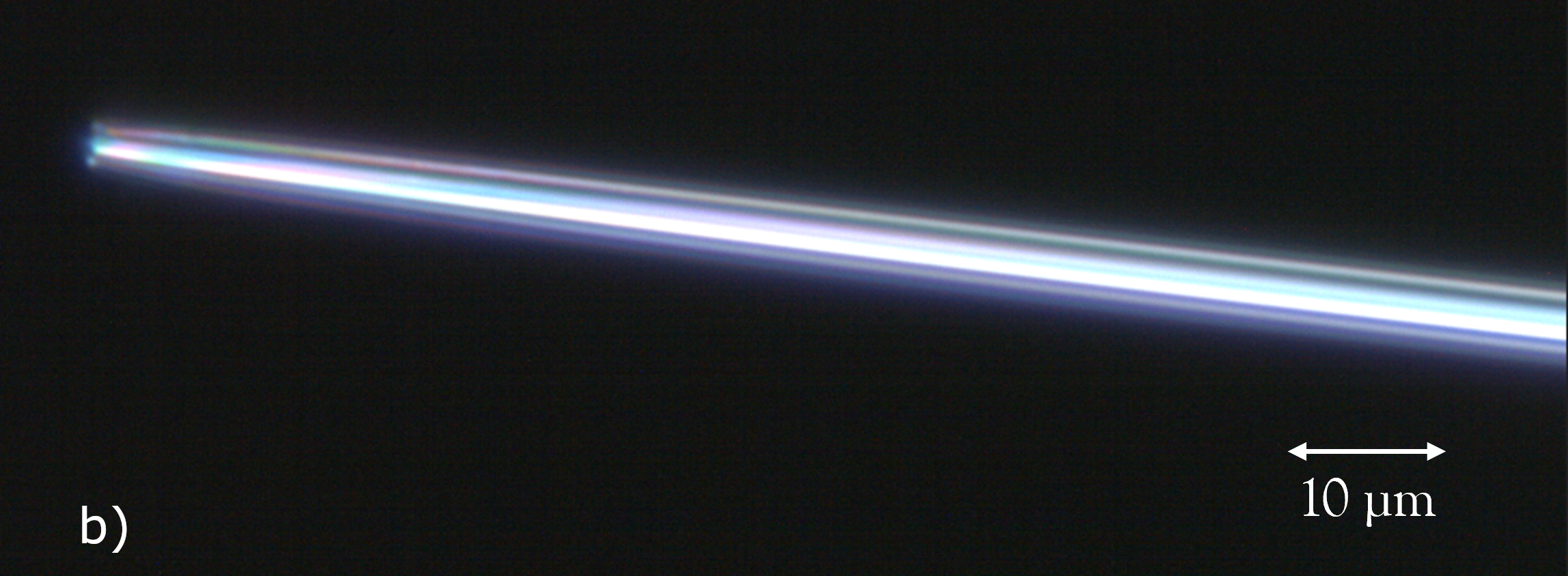}
	\caption{Representative images of laser-pulled borosilicate glass capillaries prior to probe assembly. 
		(a) and (b) show two different capillaries in the thinned region, illustrating the reproducibility of the pulling process. 
		The capillaries are fabricated from initial tubing (outer diameter 1 mm, inner diameter 0.5 mm) and stretched using a laser puller, resulting in a uniform reduction of both outer and inner diameters down to approximately 5 $\mu$m and 3 $\mu$m, respectively, over a length of several centimeters. 
		The smooth and defect-free surface, as well as the axial symmetry of the capillaries, are essential to ensure well-defined hydrodynamic boundary conditions during measurements.
	}
	\label{fig:capillaire}
\end{figure}

Representative images of the thinned capillaries are shown in Figures \ref{fig:capillaire}(a-b), illustrating the reproducibility of the pulling process and the uniformity of the capillary geometry.

\subsubsection*{Probe assembly}

A $\sim 3$ mm segment of the capillary is glued perpendicularly to one prong of the tuning fork using a UV-curable adhesive (NOA 81), with the help of micromanipulators. The stretched extremity is oriented outward and defines the probe apex.

The microsphere attachment is performed using the AFM setup equipped with an inverted optical microscope (Nikon TE2000-U), allowing bottom-view observation. The probe is vertically displaced using the piezoelectric positioner to precisely control each step of the process.

First, the capillary apex is brought into contact with a small droplet of adhesive deposited on a glass slide, allowing a minimal amount of glue to be picked up. The probe is then lowered onto a dispersion of silica microspheres (Sigma-Aldrich, nominal diameter 5 $\mu$m, ref.\ 44054-5ML-F), and a single bead is captured at the capillary apex by capillary forces. This mechanism strongly limits lateral misalignment of the microsphere and ensures an axisymmetric geometry, thereby reducing possible contributions from parasitic shear forces.

The adhesive is then cured using a UV lamp, ensuring the fixation of the microsphere at the probe apex. The amount of glue is systematically minimized in order to preserve both the cylindrical geometry of the capillary and the spherical geometry of the bead, and to avoid any perturbation of the flow field in the vicinity of the interface.

Probes exhibiting geometric defects (misalignment of the bead, excess glue, surface irregularities) are systematically discarded after inspection using an optical microscope (typical magnification $\times 50$). Only probes with a well-defined axisymmetric geometry and a clean bead-capillary junction are retained, ensuring the reproducibility of the measurements.

\subsubsection*{Geometrical characterization}

\begin{figure}[ht]
	\centering
	\includegraphics[width=0.32\textwidth]{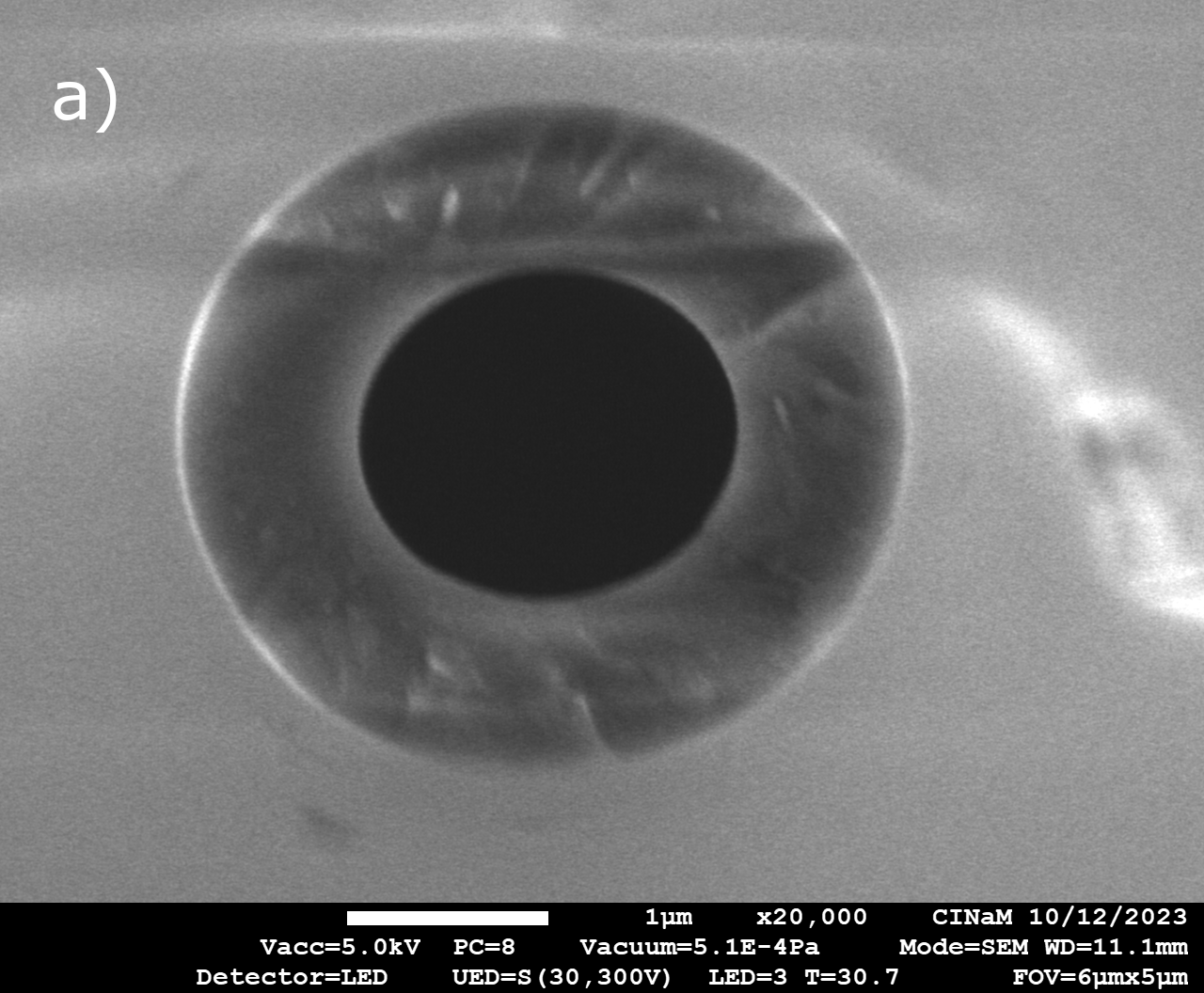}
	\includegraphics[width=0.32\textwidth]{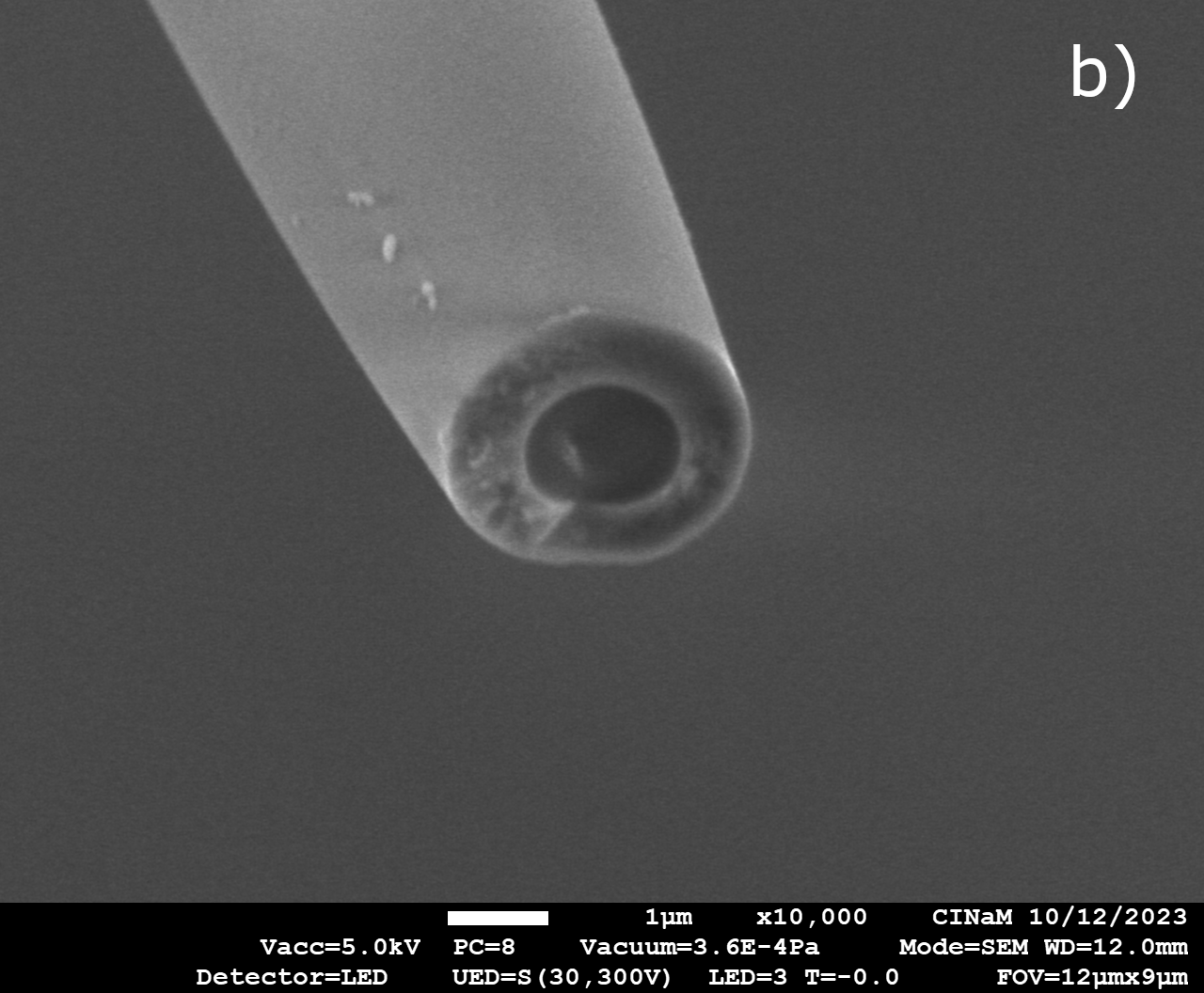}
	\includegraphics[width=0.32\textwidth]{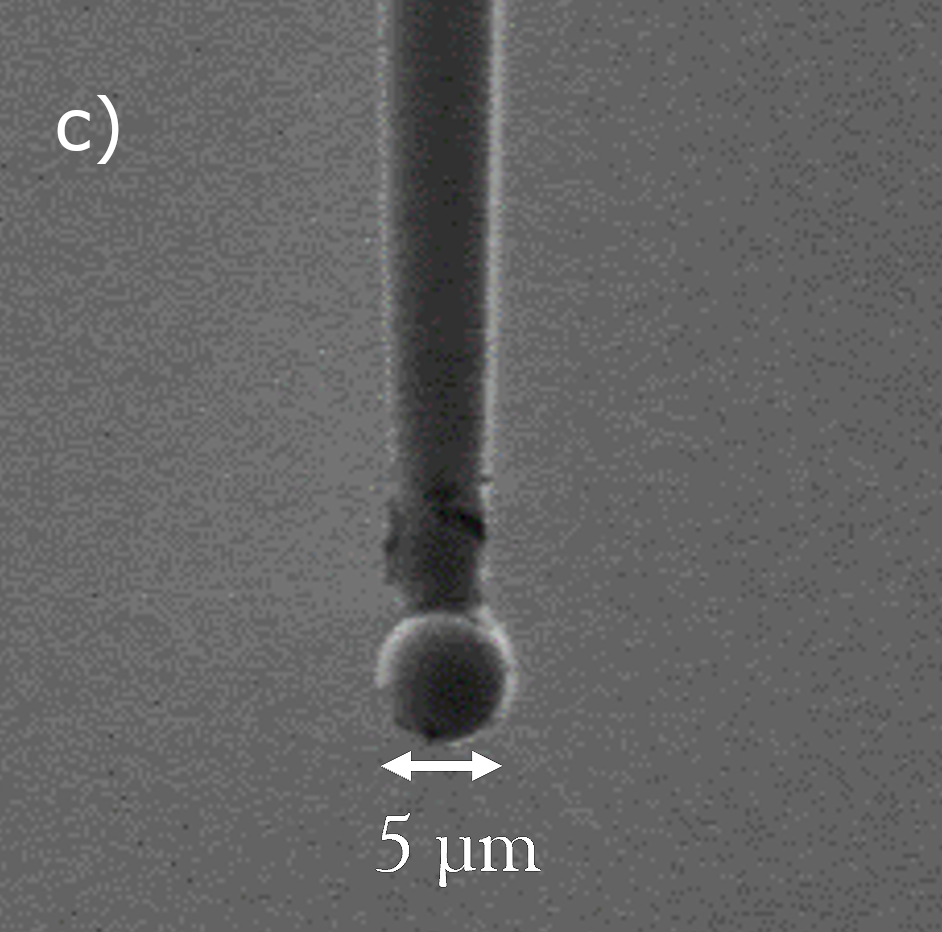}
	
	\caption{
		Scanning electron microscopy (SEM) characterization of the probe geometry. 
		(a) Side view of a laser-pulled capillary prior to bead attachment, showing the cylindrical geometry and the tapered region near the apex. 
		(b) Front view of the capillary apex, highlighting the circular cross-section and the partial constriction of the inner channel over a few microns. This localized feature remains much smaller than the typical probe-interface separation and does not significantly affect the hydrodynamic flow. 
		(c) Final probe geometry after assembly, showing a glass microsphere (diameter $\sim$ 5 $\mu$m) attached at the capillary apex using UV-curable adhesive. The bead is well-centered on the capillary axis, ensuring an axisymmetric geometry suitable for quantitative hydrodynamic measurements. 
		These observations confirm the reproducibility of the probe fabrication and the absence of significant geometric defects that could introduce measurement artifacts.
	}
	\label{fig:capillaire_et_bille}
\end{figure}

Representative scanning electron microscopy (SEM) images are shown in Figure \ref{fig:capillaire_et_bille}. Figures \ref{fig:capillaire_et_bille}(a-b) display the capillary before bead attachment (side and front views), highlighting the cylindrical geometry and the apex structure. Figure \ref{fig:capillaire_et_bille}(c) shows the complete probe after assembly, including the microsphere attached at the capillary tip.

These observations confirm the good circularity of the bead, its centering at the capillary apex, and the absence of major geometric defects likely to affect the measurements.

Overall, these experimental precautions ensure that geometric artifacts are minimized and that the measured variations in mechanical impedance primarily reflect the intrinsic response of the probed system.

\subsection{QTF calibration and operation}

\subsubsection{Stiffness calibration}

\begin{figure}[ht]
	\centering
	\includegraphics[width=0.6\textwidth]{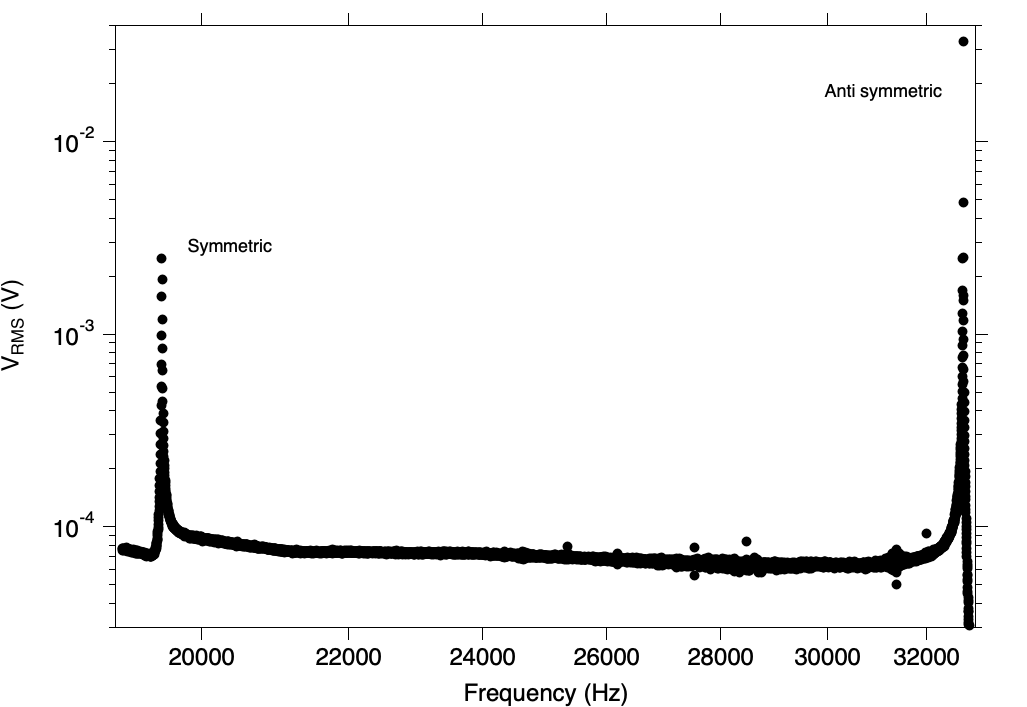}
	\caption{Plot of the RMS value of the oscillation amplitude vs excitation frequency of a sealed QTF. The signal is BP filtered and amplified (G=20dB) before being fed to the lock-in amplifier. We measure the frequency of symmetric mode $f_s=19492$Hz, and anti-symmetric (fundamental) mode $f_0=32765$Hz. Given the geometry of the tuning fork, these frequencies enable us to calculate the static stiffness constants of an arm, as well as the coupling constant, without the need for adjustable parameters. }
	\label{fig:fig1_SI}
\end{figure}

To determine the effective stiffness of the tuning fork accurately, we use the method which considers the tuning fork to be two coupled oscillators \cite{castellanos2009}. 
The design of the measuring electrodes deposited on the arms normally eliminates the contribution of the symmetric oscillation mode of the two arms. However, under gentle excitation, the slight asymmetry of the electrodes makes it possible to observe this, as shown in the figure \ref{fig:fig1_SI}.

In a system with two coupled oscillators, the two eigenmode frequencies, symmetric $f_s$ and anti-symmetric (fundamental) $f_0$ are written as

$$f_s = \frac{1}{2\pi}\sqrt{ \frac{k}{m}}$$
$$f_0 = \frac{1}{2\pi}\sqrt{ \frac{k+2k_c}{m}}$$

$k$ and $m$ are respectively the static stiffness and mass of one arm, $k_c$ the coupling stiffness. 
$m$ is determined by the geometry of a QTF's arm. Then, the measure of the eigenmode frequencies give access to the stiffnesses :

$$k= 4\pi ^2 f_s^2 m $$
$$k_c=\frac{k}{2} \left[ \left( \frac{f_0}{f_s}\right)^2-1\right] $$
and finally to the effective stiffness of the QTF

$$k_{eff} = k + 2k_c $$

$k_{eff}=4.5\pm0.4 \times 10^4$ N.m$^{-1}$ for the QTF of figure \ref{fig:fig1_SI}.

$k_{eff}$ is determined with an uncertainty of 8\%, arising mainly from the determination of the effective mass of the oscillator. This uncertainty is the main source of error in the determination of mechanical impedances.

\subsubsection{Piezoelectric response}

\begin{figure}[ht]
	\centering
	\includegraphics[width=0.6\textwidth]{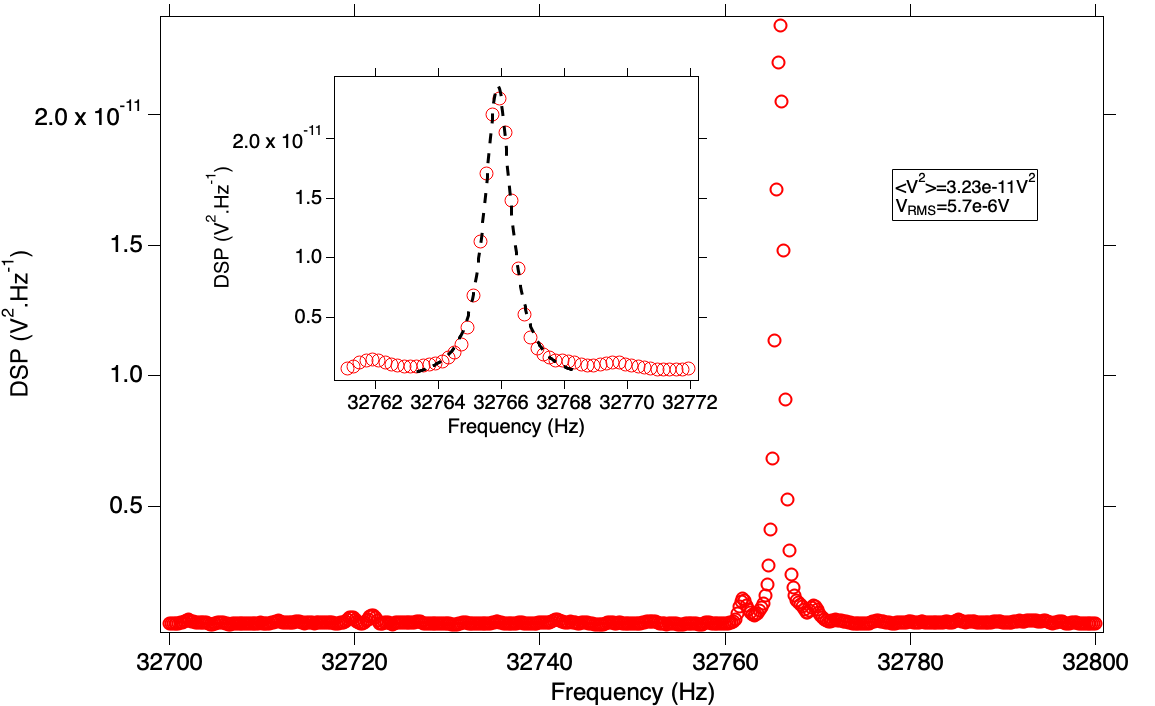}
	\caption{Estimation of the Power spectral density of the thermal noise oscillation amplitude of a sealed QTF vs frequency. Signal is BP filtered and amplified (G=40dB) before analysis. PSD is obtained from the Fourier transform of the autocorrelation function of the QTF oscillation signal following the Wiener-Khinchin theorem. At the resonance frequency, we measure the power $<V^2>=3.2\times10^{-11}$ V$^2$. This value allows us, given the stiffness of the QTF and the temperature, to determine the electrical sensitivity of the tuning fork using the equipartition theorem}
	\label{fig:fig2_SI}
\end{figure}

The piezoelectric response to oscillation is determined from the amplitude of the QTF’s thermal oscillation, as shown in figure \ref{fig:fig2_SI}. The Power Spectral Density plot is the average of 200 spectra with a resolution of 0.2Hz, from 16-bit samplings at a rate of 800 kHz. 

As with all measurements, the signal is band-pass filtered (10-100kHz pass band, with 6dB roll-off below and above cutoff frequencies) and then amplified (here G=40dB) before being processed.

Using the equipartition theorem \cite{Welker2011}, we can write the balance between the thermal energy and the elastic energy stored in the two arms of the QTF :

$$\frac{1}{2}k_b T = k_{eff}.<z>^2 $$

with $z$ the oscillation amplitude, $k_B$ the Boltzmann constant, and $T$ the temperature.

By setting $z =\alpha V$, $V$ being the piezoelectric response voltage, we can then determine the value of the electrical sensitivity, and thus the oscillation amplitude, with an uncertainty estimated to 10\%, arising from mechanical noise during measurements and numerical errors in PSD processing. We obtain here $\alpha=65.3\pm 6.5$ nm.V$^{-1}$.

\subsubsection{Harmonic oscillation of the probe}

The protocol for independently measuring the elastic and viscous components of the mechanical response is based on the assumption of a harmonic response from the probe. 

\begin{figure}[ht]
	\centering
	\includegraphics[width=0.6\textwidth]{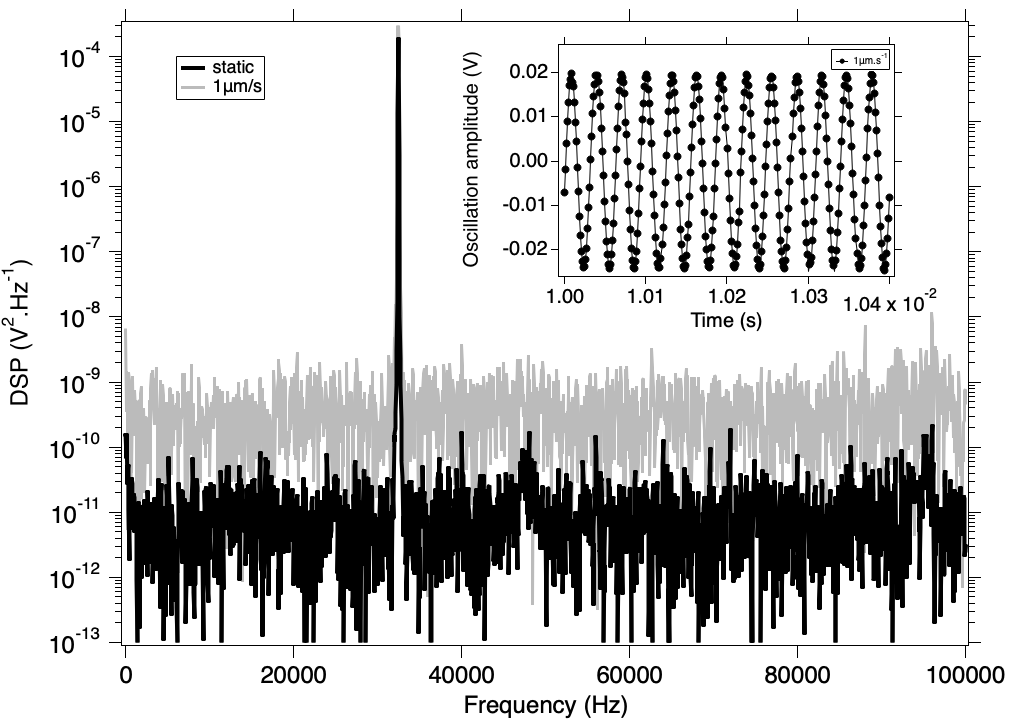}
	\caption{Power spectral density (resolution 0.4Hz) of the filtered and amplified oscillation signal of a fiber tuning fork, where the $\approx 5 \mu$m diameter fiber, which ends in a 5 $\mu$m diameter microsphere, oscillates in PDMS oil with a viscosity of 20 cSt. Oscillation amplitude $A_{RMS}=0.9$nm. Black : stationary probe. Gray : probe descending vertically through the liquid, vertical velocity $v=1\mu$m.s$^{-1}$. Whether in motion or at rest, no harmonic distortion is detectable above the noise level. Inset : Sample of the temporal signal used to process the PSD. 800kHz sampling frequency}
	\label{fig:fig3_SI}
\end{figure}

The excitation/detection chain of the force probe has been developed to optimise the signal-to-noise ratio and enables us to work directly on the temporal component of the signal of a feedback probe. Figure \ref{fig:fig3_SI}, shows such a signal recorded while the feedback probe oscillates in liquid PDMS, associated with its power spectral density.

Whether in motion or at rest, no harmonic distortion is detectable above the noise level. We have thus ensured, within the limits of our measurement resolution, that the oscillation of the feedback probe remains harmonic in a liquid medium for vertical displacement speeds of up to 1$\mu$m.s$^{-1}$.

\subsection{Reproductibility and artifacts}

\subsubsection{$D_c$ estimation}

The method for determining the position of the undeformed interface, as described in the article, is based on a linear regression.
The variance of the parameters obtained from the regression allows us to estimate the uncertainty in determining the position of the hydrodynamic zero at 10.5\% (worst case). The subsequent determination of Dc is subject to the same uncertainty. In addition, the estimation of $D_c$ depends on the choice of the fitting range used to define the linear hydrodynamic regime. Variations in this fitting window can lead to deviations in the extracted $D_c$ values of the order of $\sim 10\%$, reflecting the sensitivity of the method to the selection of the far-field region. To further assess the robustness of the data analysis, the full data processing procedure was independently implemented by two different operators using separate analysis codes. The resulting values of $D_c$ and impedance components were found to be consistent within a relative deviation of less than 5\%.

\subsubsection{Reproductibility}
Given the complexity of the probe’s manufacture, and in particular the positioning of the microsphere at its apex, the question of the reproducibility of the measurements naturally arises.

We repeated the experiment at the liquid interface to answer this question. Analysis of eight separate experiments carried out in the same conditions, using three different probes over a period of several months yields a mean value $D_c = 1.5 \pm 0.3 \mu$m.

\section*{Acknowledgements}

L.C, M.L and H.K. thank the financial support from the Centre National d'Etudes Spatiales (CNES), in the framework APR 2026 MSBioS.

\bibliography{biblio.bib}

\end{document}